\documentclass[amssymb,prl,%preprint,
twocolumn,citeautoscript,floatfix,%endfloats,
footinbib,superscriptaddress]{revtex4}
\usepackage{graphicx}
\usepackage[latin1]{inputenc}
\usepackage{times}
\usepackage{mathptm}
\newcommand\beq{\begin{equation}}
\newcommand\eeq{\end{equation}}
\newcommand\bea{\begin{eqnarray}}
\newcommand\eea{\end{eqnarray}}

\newcommand\ben{\begin{enumerate}}
\newcommand\een{\end{enumerate}}

\newcommand\LMO{Li$_{0.9}$Mo$_6$O$_{17}$}

\begin{document}

\title{Superconducting critical field far above the Pauli limit in one-dimensional Li$_{0.9}$Mo$_6$O$_{17}$}

\author{J.-F. Mercure}
\affiliation{H. H. Wills Physics Laboratory, University of Bristol, Tyndall Avenue, BS8 1TL, United Kingdom}

\author{A. F. Bangura}
\affiliation{H. H. Wills Physics Laboratory, University of Bristol, Tyndall Avenue, BS8 1TL, United Kingdom}
\affiliation{RIKEN (The Institute of Physical and Chemical Research), Wako, Saitama, 351-0198, Japan}

\author{Xiaofeng Xu}
\affiliation{H. H. Wills Physics Laboratory, University of Bristol, Tyndall Avenue, BS8 1TL, United Kingdom}
\affiliation{Department of Physics, Hangzhou Normal University, Hangzhou 310036, China}

\author{N. Wakeham}
\affiliation{H. H. Wills Physics Laboratory, University of Bristol, Tyndall Avenue, BS8 1TL, United Kingdom}

\author{A. Carrington}
\affiliation{H. H. Wills Physics Laboratory, University of Bristol, Tyndall Avenue, BS8 1TL, United Kingdom}

\author{P. Walmsley}
\affiliation{H. H. Wills Physics Laboratory, University of Bristol, Tyndall Avenue, BS8 1TL, United Kingdom}

\author{M. Greenblatt}
\affiliation{Department of Chemistry and Chemical Biology, Rutgers University, Piscataway, NJ 08854}

\author{N. E. Hussey}
\affiliation{H. H. Wills Physics Laboratory, University of Bristol, Tyndall Avenue, BS8 1TL, United Kingdom}

\begin{abstract}
The upper critical field $H_{c2}$ of purple bronze \LMO~is found to exhibit a large anisotropy, in quantitative agreement with that expected from the observed electrical resistivity anisotropy. With the field aligned along the most conducting axis, $H_{c2}$ increases monotonically with decreasing temperature to a value five times larger than the estimated paramagnetic pair-breaking field. Theories for the enhancement of $H_{c2}$ invoking spin-orbit scattering or strong-coupling superconductivity are shown to be inadequate in explaining the observed behavior, suggesting that the pairing state in \LMO\ is unconventional and possibly spin-triplet.
\end{abstract}

\maketitle

Superconductivity in quasi-one-dimensional (q1D) conductors has attracted sustained interest from the theoretical community \cite{Emery86}, largely due to the fact that under certain conditions, rare phenomena such as spin-triplet pairing \cite{Podolsky04, Nickel05, Kuroki05} or the spin-singlet, spatially inhomogeneous Fulde-Ferrell-Larkin-Ovchinnikov (FFLO) state \cite{Buzdin83, Lebed86, Dupuis93} may be realized. The organic conductors (TMTSF)$_2$X (X = PF$_6$, ClO$_4$) have been most extensively studied in this regard, though the nature of their pairing state has not yet been fully determined. In (TMTSF)$_2$PF$_6$, the constant Knight shift across the superconducting (SC) transition $T_c$ \cite{Lee02a}, together with the observed violation of the Pauli paramagnetic limit \cite{Lee00} supports triplet pairing, while in (TMTSF)$_2$ClO$_4$, a Knight shift suppression below $T_c$ and the presence of line nodes \cite{Shinagawa07} indicate collectively a $d$-wave, spin-singlet pairing state. The possible realization of the FFLO state in (TMTSF)$_2$ClO$_4$ at low $T$, as suggested by recent angular studies of $H_{c2}$ \cite{Yonezawa08}, is also consistent with singlet rather than triplet pairing. Theoretically, the coexistence of spin-(SDW) and charge-(CDW) density-wave instabilities can lead to a complex phase diagram where both singlet and triplet phases lie in close proximity, with triplet pairing becoming dominant as repulsive interchain interactions are enhanced \cite{Kuroki01, Fuseya05, Nickel05}. Intriguingly, the triplet state known to exist in the q2D perovskite superconductor Sr$_2$RuO$_4$ might also arise from repulsive interactions between q1D bands \cite{ Raghu10}.

\LMO\ (LiMO) is a transition metal oxide with q1D electronic properties. It is metallic at high $T$, semiconducting below a temperature 15 K $\leq T_{\rm min} \leq$ 30 K and
superconducting below $T_c \sim$ 2 K \cite{Greenblatt84}. Despite having a $T_c$ higher than the (TMTSF)$_2$X family, its SC properties have received little attention to date. While the presence of a density wave (DW) transition in LiMO was initially discounted, recent magnetotransport data appear to suggest some form of DW gapping \cite{Xu09}. The precise nature of the DW however, and its relation to the superconductivity, has yet to be resolved. Finally, signatures of superconductivity have been found to emerge at high magnetic fields \cite{Xu09} in LiMoO crystals that are {\it non}-superconducting in zero-field, consistent with theoretical predictions for a q1D superconductor with triplet pairing \cite{Lebed86, Dupuis93}.

\begin{figure}[hbt]
  \centering
\includegraphics[width=7.5cm,keepaspectratio=true]{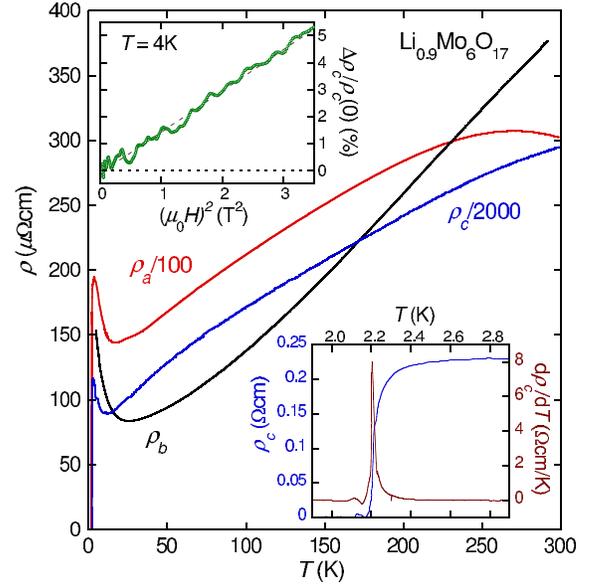}
\caption{Zero-field resistivity curves for \LMO\ for {\bf I}$\| a,b$ and $c$, scaled by 2000 ($\rho_c$) and 100 ($\rho_a$) for clarity. Note that the $\rho_b(T)$  trace only goes down to 4.2 K, as  explained in the Supplementary Information.  Lower inset: Superconducting transition as seen in $\rho_c(T)$ (blue) and its temperature derivative d$\rho_c$/d$T$ (brown). Upper inset: $c$-axis magnetoresistance $\Delta \rho_c / \rho_c$ on the same single crystal at $T$ = 4 K, i.e. just above $T_c$, plotted versus $H^2$ ({\bf H}$\| a$). The slope gives a measure of the in-chain mean-free-path (see text).}
	\label{fig: TRampInset2}
\end{figure}

\begin{figure*}[t]
  \centering
	\includegraphics[width=15.0cm,keepaspectratio=true]{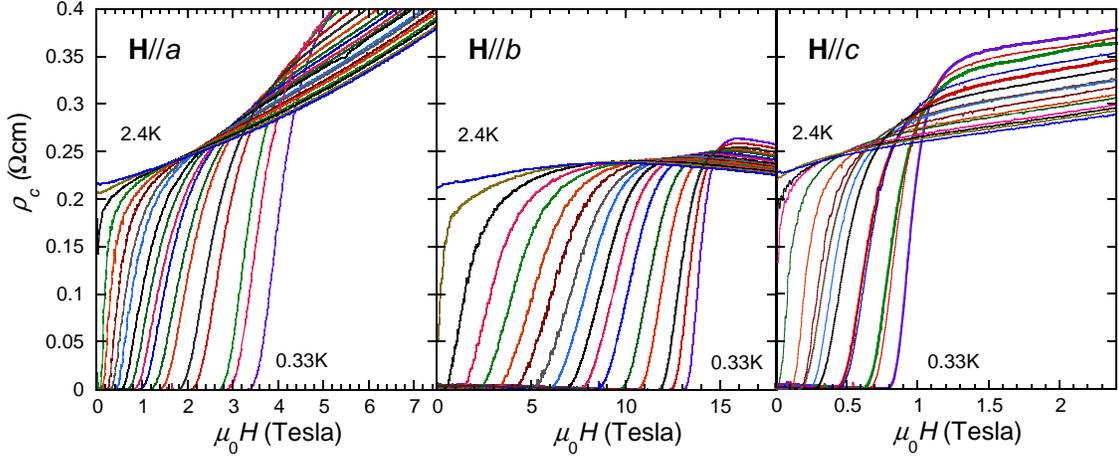}
	\caption{Field sweeps of the $c$-axis resistivity of Li$_{0.9}$Mo$_6$O$_{17}$ in $\sim$ 0.1 K steps for {\bf H} aligned along the three crystallographic axes.}
	\label{fig: AllTransitions2}
\end{figure*}

Here, we report a detailed temperature and orientational study of $H_{c2}$ in crystals that are superconducting in zero-field and extract $H_{c2}(T)$ for fields applied along the three crystallographic axes. With the field parallel to the zigzag chains ({\bf H}$ \| b$), $H_{c2}$ increases monotonically with decreasing temperature to a value five times larger than the usual Pauli paramagnetic limit. We show evidence that LiMoO is a strongly-coupled superconductor in the clean limit. However, the large $H_{c2}$ values can neither be explained wholly by the effects of strong coupling \cite{Carbotte90} nor by spin-orbit scattering as parameterized by the Werthamer-Helfand-Hohenberg (WHH) theory \cite{WHH66}. Such a finding points to the possibility that the low-field SC state in LiMO might also be a spin-triplet.

Figure~\ref{fig: TRampInset2} presents zero-field resistivity $\rho(T)$ curves from 300~K down to 1.6~K (4.2~K) for {\bf I}$\| a, c$ ({\bf I}$\| b$) respectively, scaled to incorporate all three curves on the same set of axes. (Full details of the samples and measurement techniques are given in the Supplementary Information.) On cooling from room temperature, $\rho(T)$ is metallic down to about 15 K, followed by a well-defined upturn and finally, superconductivity. The size of the resistivity upturn is much smaller than found in the (non-superconducting) crystals described in Ref.~\cite{Xu09}, consistent with the anti-correlation between $T_c$ and the size of the resistivity upturn first reported by Matsuda {\it et al.} \cite{Matsuda86}. The lower inset shows a blow-up of the $c$-axis resistive transition at $T_c$, the mid-point of which, as defined using the maximum in d$\rho_c$/d$T$, occurs at 2.2~K. The resistivity anisotropy extracted from this set of curves is 80:1:1600 (150:1:1600) for $\rho_a$:$\rho_b$:$\rho_c$ at $T$ = 300 K (4.2 K) respectively. This is almost two orders of magnitude larger than the anisotropies reported in the recent literature \cite{daLuz07, Chen10}, highlighting the extreme care needed to isolate the individual current directions in such low-dimensional systems (see the SI section for further details).

Figure~\ref{fig: AllTransitions2} shows $c$-axis resistivity curves obtained in an 18 Tesla pumped $^3$He cryostat with the field applied along $a$, $b$, and $c$, for temperatures between 0.33 and 2.40~K. The $\rho_c(H)$ curves for {\bf H}$\| b$ show an unusual broadening at intermediate temperatures, the origin of which is not understood at present. The large anisotropy in $H_{c2}$ is apparent from inspection of the field scales in the three different panels. Field alignment for {\bf H}$\| b$ required an accuracy $<$ 1$^{\circ}$ that was difficult to achieve in our $^3$He system and higher $H_{c2}$ values ({\bf H}$\| b$) were observed in a second set of measurements performed on the same crystal in a pumped $^4$He system. The phase diagram for $H_{c2}(T)$ obtained from fixed-field temperature sweeps in the latter is shown in Fig.~\ref{fig: Gradients2}, where the open squares, two-tone squares and open circles refer to measurements performed with {\bf H}$\| a,b,c$ respectively. Here, $H_{c2}(T)$ is determined by the maximum in the relevant derivative d$\rho_c$/d$T$ or d$\rho_c$/d$H$. As shown in the SI section, choosing a different criterion does not change qualitatively the overall behavior, nor the anisotropy parameters. All data in Figs.~\ref{fig: AllTransitions2} and ~\ref{fig: Gradients2} were corrected for the remnant field by symmetrizing with respect to positive and negative field values. Data for a second single crystal with a slightly lower $T_c$ and $H_{c2}$(0) are shown (for {\bf H}$\| b$ only) in the SI section.

From the initial slopes of the increase in $H_{c2}(T)$ below $T_c$ (dashed lines in Fig. \ref{fig: Gradients2}), we obtain d$H_{c2}$/d$T|_{T \leq T_c}$ values of -1.5~T/K, -19.5~T/K and -0.5~T/K and a corresponding critical field ratio of 3:39:1 for {\bf H}$\| a$:$b$:$c$ respectively. According to anisotropic Ginzburg-Landau (GL) theory
\beq
{H_{c2}^i \over H_{c2}^j} = {\xi_i \over \xi_j} = {\sqrt{\sigma_{ii}}\over \sqrt{\sigma_{jj}}} = {\sqrt{\rho_{jj}}\over \sqrt{\rho_{ii}}},
\label{eq:ratios}
\eeq
where subscripts $i,j$ refer to crystalline axes, superscripts $i,j$ refer to the direction of the applied magnetic field, $\xi_{i,j}$ ($\propto v_{i,j}$) is the orientation dependent coherence length, $\sigma_{ii}$/$\sigma_{jj}$ ($\propto v_{ii}^2/v_{jj}^2$) is the anisotropy in the diagonal elements of the conductivity tensor and $v_{i,j}$ are the respective Fermi velocities. Squared, the SC anisotropy is approximately 9:1500:1, or 170:1:1500 when inverted. Note that these ratios are in excellent quantitative agreement with those (150:1:1600) obtained from the normal state resistivity measurements.

A more complete $H_{c2}(T)$ phase diagram, extracted from the data plotted in Fig. \ref{fig: AllTransitions2}, is shown in Fig. \ref{fig: PhaseDiag}. (Note that the $H_{c2}$  values for
{\bf H}$\| a,c$ have been re-scaled in this plot). Using the (linearly) extrapolated zero-temperature values $H_{c2}(0)$ from the phase diagram and the equation
\beq
H_{c2}^i(0) = {\Phi_0 \over 2 \pi \xi_j(0) \xi_k(0)},
\label{eq:Hc2}
\eeq
we obtain estimates for the three coherence lengths; $\xi_b$(0) $\simeq$ 300 ${\rm \AA}$, $\xi_a$(0) $\simeq$ 100 ${\rm \AA}$ and $\xi_c$(0) $\simeq$ 25 ${\rm \AA}$. Significantly, the interchain coherence lengths are both longer than the lattice spacing (or more precisely 2$\xi_{a(c)} (0) > a(c) - d$ \cite{Hussey96}, where $a = 12.73 {\rm \AA}$ and $c = 9.51 {\rm \AA}$ are the $a, c$-axis lattice constants and $d \sim 3 {\rm \AA}$ is the approximate width of the MoO$_4$ octahedra), implying the absence of Josephson coupling and a continuous phase of the SC order parameter across the chains. Hence, despite the extreme one-dimensionality of LiMO in the normal state, its superconductivity appears to be described satisfactorily using anisotropic-3D GL theory.

\begin{figure}[t]
  \centering
	\includegraphics[width=7.5cm,keepaspectratio=true]{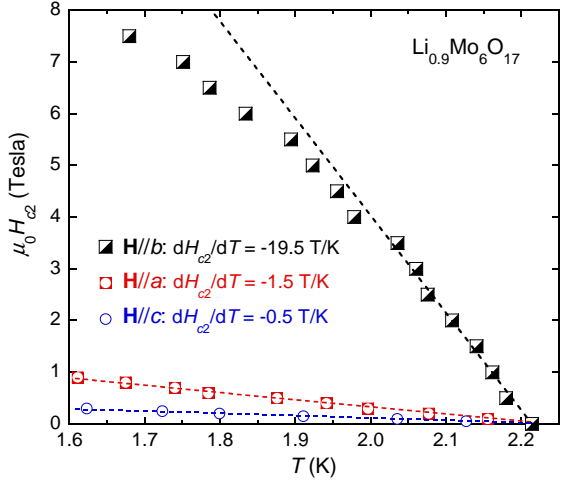}
	\caption{Phase diagram for $H_{c2}$ as a function of temperature for fields applied along the three crystallographic directions and temperatures down to 1.6~K obtained with the $^4$He vapor pressure system.}
	\label{fig: Gradients2}
\end{figure}

For a q1D spin-singlet superconductor with {\bf H}$\|$chain, orbital pair-breaking is minimized (due to the small interchain electron velocities) and superconductivity can only be destroyed once the Zeeman energy arising from spin-splitting of Cooper pairs exceeds the SC condensation energy. For an isotropic BCS superconductor, the Pauli limit is expressed as $\mu_0 H_P = 1.84 T_c \simeq$ 4.0~T for $T_c$~=~2.2~K. $H_P$ can also be calculated independently \cite{Zuo00} using actual values (see SI section) for the Pauli susceptibility $\chi_p$ (= 2.8 x 10$^{-6}$) and for the condensation energy $U_c$ (= 2.2 mJ/mol, estimated from the specific heat anomaly at $T_c$) obtained on crystals taken from the same batch and with similar $T_c$ values. These give $\mu_0 H_p \simeq 3.1$~T for LiMO, i.e. comparable with the BCS value but still five times smaller than the measured $H_{c2}$(0) for {\bf H}$\| b$.

According to WHH theory \cite{WHH66}, spin-orbit scattering (e.g. at the Mo site) can act to limit Zeeman splitting and thus increase the value of $H_{c2}$ beyond the usual Pauli limit. It is expressed using two dimensionless parameters, the Maki parameter $\alpha$ and the spin-orbit scattering $\lambda_{\rm SO}$. The former is constrained through the expression $\alpha \simeq 0.528$ d$H_{c2}$/d$T|_{T \leq T_c}$ (= 7.5 for the $^3$He data), while $\lambda_{\rm SO}$ = 2$\hbar / 3 \pi k_B T_c \tau_{\rm SO}$ (= 32) is determined from the best curve fit to the data, shown in Fig. \ref{fig: PhaseDiag} by a dashed line. The associated spin orbit scattering time $\tau_{\rm SO}$ can be converted to a mean-free-path $\ell_{\rm SO}$ using the measured (in-chain) Fermi velocity \cite{Wang06}. For both data sets (i.e. from the $^3$He (Fig.~\ref{fig: PhaseDiag}) and $^4$He cryostats (Fig.~\ref{fig: Gradients2})), $\ell_{\rm SO} \sim$ 120 ${\rm \AA}$.

\begin{figure}[t]
  \centering
	\includegraphics[width=7.5cm,keepaspectratio=true]{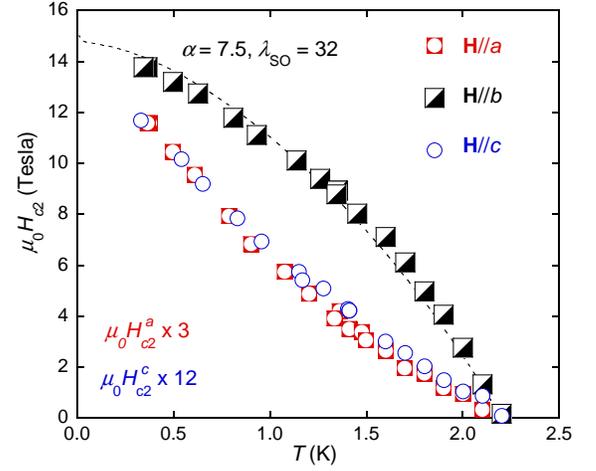}
	\caption{Phase diagram for $H_{c2}(T)$ for temperatures down to 0.3~K obtained with the $^3$He cryostat. The data were obtained from the field derivative of the magnetic field sweep curves shown in Fig.~\ref{fig: AllTransitions2}. Note that for {\bf H}$\| a,c$, the $H_{c2}$ values have been normalized to emphasize the similarities of their $T$-dependencies. The dashed line is a fit to the WHH theory with $\alpha$ = 7.5 and $\lambda_{\rm SO}$ = 32 (see text for details).}
	\label{fig: PhaseDiag}
\end{figure}

An estimate for the transport mean-free-path $\ell_0$ can be obtained from the low-$T$ interchain magnetoresistance (MR). With {\bf I}$\| c$ and {\bf H}$\| a$, Boltzmann theory gives for a q1D metal in the weak-field limit
\beq
{\Delta \rho_c \over \rho_c} = \bigg({ecB \over \hbar}\bigg)^2 \ell_0^2,
\label{eq:meanfreepath}
\eeq

As shown in the top inset of Fig. \ref{fig: TRampInset2}, $\Delta \rho_c$/$\rho_c \propto B^2$ at low fields with a slope that yields $\ell_0 \sim$ 650 ${\rm \AA}$ at $T$ = 4 K. According to the Abrikosov formula \cite{Abrikosov}, $\tau_0$/$\tau_{\rm SO}$ = $(Z/137)^4$ ($Z$ = 42 for Molydenum). Thus, one expects $\tau_{\rm SO}$ to be $\sim$ 100 times longer than the transport lifetime $\tau_0$, in marked contrast to what is obtained from the WHH parameters. Moreover, the diffusion constant obtained from the Maki parameter gives $\ell_0 < 1 {\rm \AA}$, showing clearly that WHH theory is inapplicable here (it is only truly valid in the dirty limit). It also shows that the spin-orbit scattering alone cannot account for the values of $H_{c2}$ observed in LiMO.

Strong (e.g. electron-phonon) coupling can also act to reduce the effects of Pauli limiting through renormalization of the band splitting. Fits to the specific-heat anomaly yield a coupling constant $\lambda$ = 1.2 $\pm$ 0.1, intermediate between those of Nb and Pb. Even with this strength of coupling however, the magnitude of $H_{c2}$(0) in LiMO is more than double the renormalized value (see SI section for more details) \cite{Carbotte90}. While one cannot exclude the possibility that the high $H_{c2}$ found in LiMO is due to a combination of singlet pairing, strong-coupling superconductivity {\it and} strong spin-orbit scattering, it would require a precise combination of all of these effects to realize the present situation.

In a triplet superconductor, the spins of the Cooper pair can be co-aligned, making the Pauli pair-breaking effect redundant. Triplet superconductivity is, however, extremely fragile and can easily be destroyed by impurities \cite{Mackenzie98}. For our samples, $\ell_0 > 2\xi_b$(0), i.e.~within the clean limit, and while the FFLO state cannot be ruled out at the lowest temperatures and highest fields, significantly we find in LiMO, three of the key ingredients for realizing q1D triplet superconductivity: extreme one-dimensionality, a $H_{c2}(T)$ profile with minimal paramagnetic limiting and a long mean-free-path. The recent observation of scaling in the longitudinal $b$-axis MR in LiMO \cite{Xu09} provides compelling evidence that some form of DW develops below $T_{\rm min}$. As discussed in the introduction, the nature of the DW fluctuations near $T_c$ may ultimately determine the pairing state. In q1D systems close to a Peierls-type CDW instability, $s$-wave is the dominant pairing channel \cite{Bakrim10}. However structural \cite{Santos07}, thermodynamic \cite{Matsuda86} and optical studies \cite{Choi04} have all failed to find evidence of a genuine phase transition in LiMO at $T = T_{\rm min}$. We also find no evidence of a specific heat anomaly at $T_{\rm min}$, contrary to an earlier report \cite{Schlenker85}. This lack of evidence has lent support to the notion that the CDW instability in LiMO is in fact driven by {\it electronic}-interactions  \cite{Santos07}, as is the case for strongly-interacting coupled Luttinger liquids \cite{Boies95}, with either singlet or triplet pairing competing for the ground state.

Finally, for {\bf H}$\perp$$b$, $H_{c2}(T)$ displays pronounced upward curvature, leading to a reduction in the SC anisotropy with decreasing temperature. A similar enhancement has been observed in (TMTSF)$_2$PF$_6$ and attributed to the formation of insulating (SDW) and SC domains at pressures approaching the SDW phase \cite{Lee02b}. This model, in which $H_{c2}$ is determined by the largest penetration depth perpendicular to the applied field, can account for the very similar $H_{c2}$ values found for {\bf H}$\| a$ and {\bf H}$\| b$ in q1D (TMTSF)$_2$PF$_6$ as well as the upturn in $H_{c2}$ for {\bf H}$\| c$. In LiMO however, there is no clear evidence for such an SDW phase at these temperatures and field scales and the anisotropy in $H_{c2}$, though reduced with decreasing temperature, always remains large.

An alternative explanation for the enhancement of $H_{c2}$ is a field-induced reduction in the effective dimensionality of the electronic system, as observed in other q1D conductors \cite{Yonezawa08, Behnia95, Lee97, Hussey02}. For {\bf H}$\perp$$b$, the semiclassical motion of quasiparticles along, say, $\vec{k}$ oscillates with an amplitude $t_k / e v_F B$,  where $t_k$ is the $\vec{k}$-axis hopping parameter. (Note that even though $\rho_c(T)$ is non-metallic at low-$T$, the positive, quadratic transverse MR shown in the inset to Fig.~\ref{fig: TRampInset2} indicates that coherent quasiparticles do exist in this temperature regime.) As $B$ increases, the amplitude of the oscillatory interchain motion decreases \cite{GorkovLebed}. This gradual confinement suppresses orbital pair-breaking and leads to an overall decrease in $v_{k(j)}$ and in turn  $\xi_k$, which now becomes field-dependent. As a result, $H_{c2}^i$ increases with decreasing temperature. While this field-induced dimensional crossover will occur in {\it both} orthogonal field orientations, it is expected to happen at different field scales \cite{Narduzzo06}. Moreover, it remains an open question whether this confinement process can lead to an enhancement of $H_{c2}$ at fields well below the crossover field. In the case of triplet pairing, this field-induced dimensional crossover can lead ultimately to re-entrant superconductivity \cite{Lebed86, Dupuis93}, though in LiMO, such behavior has only been seen to date with the field aligned {\it parallel} to the molybdate chains and only in samples that do not superconduct in zero-field \cite{Xu09}. In the present batch of crystals, we observe a strong negative magnetoresistance for {\bf H}$\| b$, but as yet, no sign of re-entrant superconductivity.

In conclusion, we have demonstrated the existence of a highly anisotropic, yet still three-dimensional SC state in LiMO with an anisotropy (near $T_c$) that is in excellent quantitative agreement with the measured normal-state electrical anisotropy. The magnitude of the $b$-axis upper critical field exceeds the usual Pauli limit by a factor of five. We have shown that neither spin-orbit scattering nor strong-coupling superconductivity can account fully for this enhanced $H_{c2}$ and thus have speculated that LiMO is a viable candidate for the realization of triplet superconductivity. Although the effective dimensionality of the electronic state just above $T_c$ is yet to be determined, LiMO displays all the hallmarks of a 1D Luttinger liquid at least above $T = T_{\rm min}$ \cite{Wang06, Wakeham11}, suggesting that the superconductivity may in fact have a higher dimensionality than the normal state out of which it condenses. In this case, pairing has to involve electrons on different chains, thus providing a test-bed for theoretical claims that triplet pairing in q1D superconductors is stabilized in the presence of (repulsive) interchain interactions \cite{Kuroki01, Fuseya05, Nickel05}.

\acknowledgements{We thank J. F. Annett, P. Chudzinski, B. Gyorffy, R. H. McKenzie, J. Merino and N. Shannon for stimulating discussions. This work was supported by the EPSRC (UK). XX acknowledges financial support from NSFC (No. 11104051), NEH a Royal Society Wolfson Research Merit Award.}

\newpage

\title{Supplementary Information for "Superconducting critical field far above the Pauli limit in one dimensional Li$_{0.9}$Mo$_6$O$_{17}$"}

\author{J.-F. Mercure}
\affiliation{H. H. Wills Physics Laboratory, University of Bristol, Tyndall Avenue, BS8 1TL, United Kingdom}

\author{A. F. Bangura}
\affiliation{H. H. Wills Physics Laboratory, University of Bristol, Tyndall Avenue, BS8 1TL, United Kingdom}
\affiliation{RIKEN (The Institute of Physical and Chemical Research), Wako, Saitama, 351-0198, Japan}

\author{Xiaofeng Xu}
\affiliation{H. H. Wills Physics Laboratory, University of Bristol, Tyndall Avenue, BS8 1TL, United Kingdom}
\affiliation{Department of Physics, Hangzhou Normal University, Hangzhou 310036, China}

\author{N. Wakeham}
\affiliation{H. H. Wills Physics Laboratory, University of Bristol, Tyndall Avenue, BS8 1TL, United Kingdom}

\author{A. Carrington}
\affiliation{H. H. Wills Physics Laboratory, University of Bristol, Tyndall Avenue, BS8 1TL, United Kingdom}

\author{P. Walmsley}
\affiliation{H. H. Wills Physics Laboratory, University of Bristol, Tyndall Avenue, BS8 1TL, United Kingdom}

\author{M. Greenblatt}
\affiliation{Department of Chemistry and Chemical Biology, Rutgers University, Piscataway, NJ 08854}

\author{N. E. Hussey}
\affiliation{H. H. Wills Physics Laboratory, University of Bristol, Tyndall Avenue, BS8 1TL, United Kingdom}

\maketitle

\section{Sample characterization and resistivity measurements}

Single crystals of \LMO~were grown using a temperature gradient flux method \cite{McCarroll84}. The crystals reported in this Letter were all selected from the same batch and face-indexed
using a single-crystal X-ray diffractometer to check their crystallinity and to identify the $a$- and $b$-axes. The lattice parameters were $a$ = 12.73 ($\pm$ 0.01)\,\AA, $b$
= 5.53 ($\pm$ 0.01)\,\AA and $c$ = 9.51 ($\pm$ 0.01)\,\AA, while the monoclinic angle ($\beta$) was determined to be 90.6$^{\rm o}$, in agreement with previous crystallographic
studies \cite{Onoda87, daLuz11}. Once face-indexed, the selected samples were then cut into rectangles of appropriate dimensions along the $a$- and $b$-axes and cleaved to obtain clean
surfaces onto which electrical contacts were made.

For the $\rho_a(\rho_c)$ measurements, we used the so-called quasi-Montgomery method in which the resistivity is measured in only one (the most resistive) configuration. In both cases,
the dimensions of the crystal were chosen such that the $a$- or $c$-axis dimension of the equivalent isotropic crystal, estimated by multiplying the true $a$- or $c$-axis dimension by
the square of the resistive anisotropy \cite{vdPauw}, was always at least one order of magnitude longer than the corresponding length along $b$. This ensured that the current flow was
uniaxial along $a$ or $c$.

For the $\rho_a(T)$ measurements, crystals were cut into squares of approximate dimensions 300 x 300 x 20 $\mu$m$^3$ and four gold wires were attached with Ag paint at the corners of each crystal, which itself was raised above the sample platform with the current/voltage contacts acting as posts in order to short out the highly resistive $c$-axis component. For the $\rho_c(T)$ measurements, thicker samples were employed, typically 200 x 200 x 50 $\mu$m$^3$ with the gold wires painted onto the faces of the crystal in such a way as to short out the more resistive $a$-axis. In both cases, the interchain resistivities were measured using ac lock-in detection techniques with excitation currents of 1 - 10 $\mu$A. No non-linear effects were observed, neither in the normal state nor in the mixed state, over this range of currents.

\begin{figure}[t]
  \centering	
    \includegraphics[width=8.0cm,keepaspectratio=true]{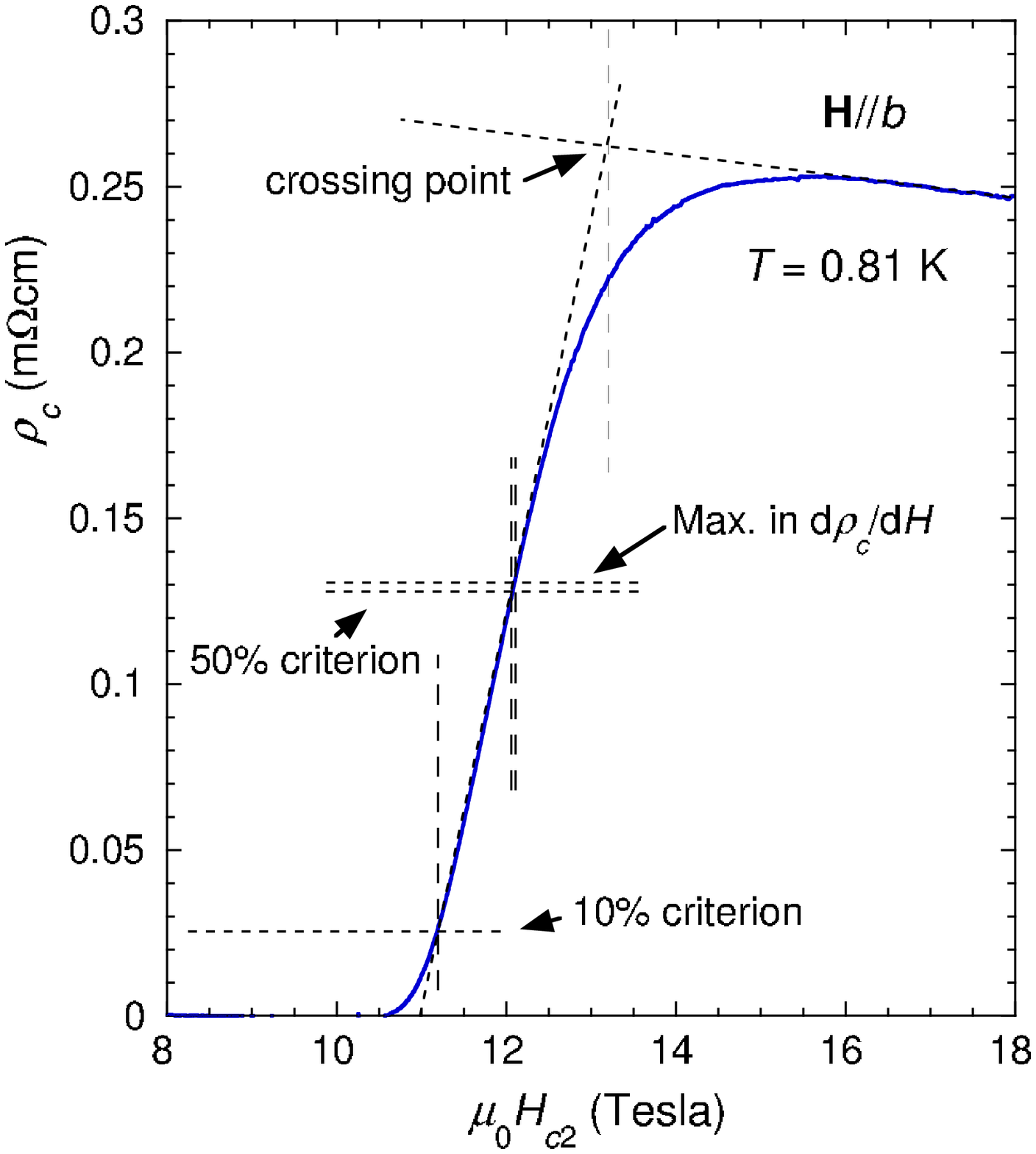}
	\caption{Demonstration of the different criteria used to determine $H_{c2}$ in \LMO. The solid blue line is $\rho_c(H)$ field-sweep data obtained at $T$ = 0.81 K with {\bf H}$\| b$.
The 10\% and 50\% criteria refer to the field scale at which the resistivity reaches 10\% and 50\% respectively of the maximum in the $\rho_c(H)$ curve. The crossing-point criterion
corresponds to the field scale at which extrapolations of the mixed-state and normal-state resistivities meet. Note that the 50\% criterion coincides with the maximum in d$\rho_c$/d$T$, as
expected.}
%	\label{S1}
\end{figure}

\begin{figure*}[t]
  \centering
	\includegraphics[width=15cm,keepaspectratio=true]{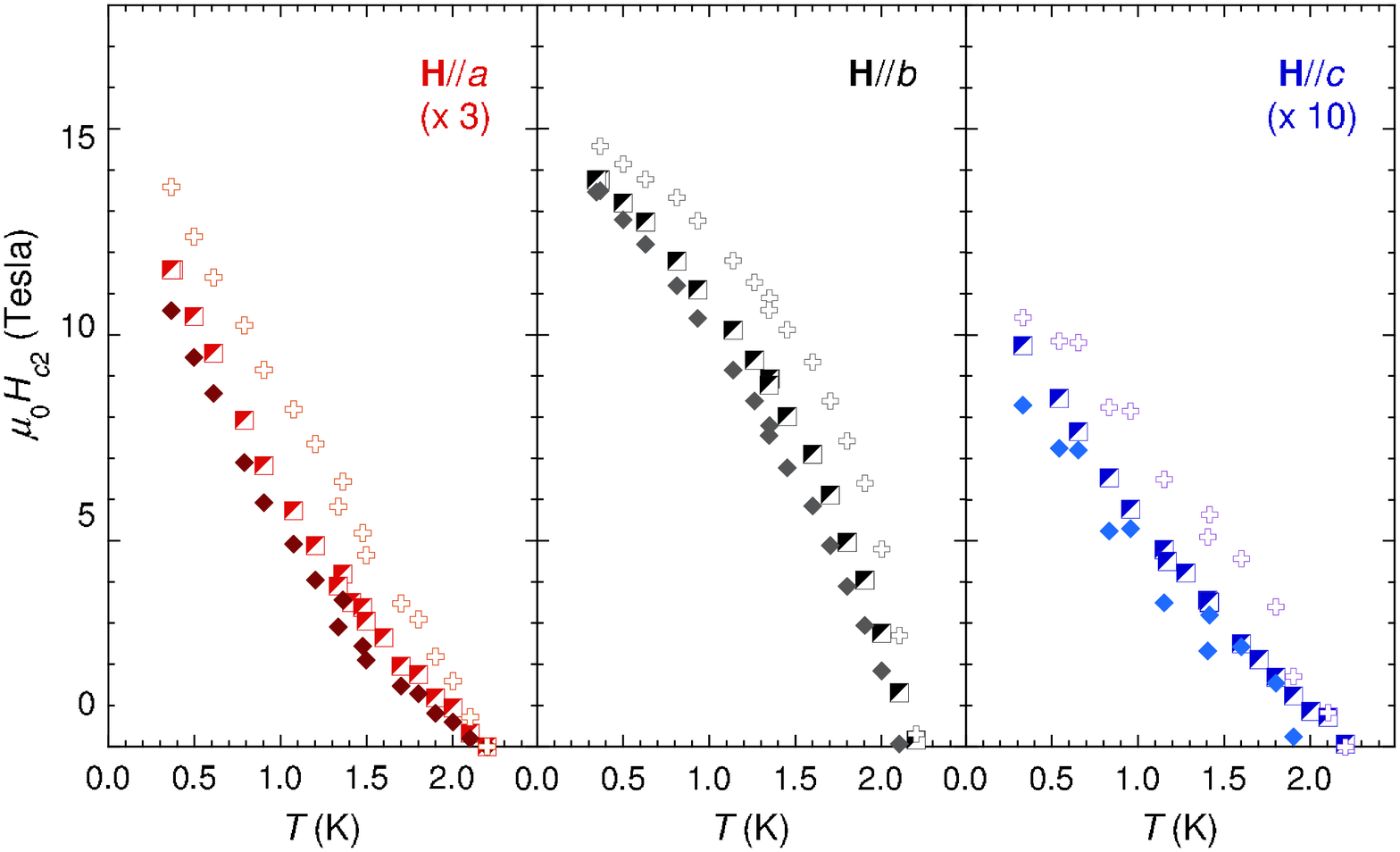}
	\caption{Comparison of the $H_{c2}(T)$ curves along {\bf H}$\| a$:$b$:$c$ for the different criteria described in Figure S1. The diamonds, two-tone squares and crosses correspond to
the 10\%, maximum in d$\rho_c$/d$T$ and the crossing-point criteria respectively.}
%	\label{S2}
\end{figure*}

Of all the individual components of the electrical conductivity tensor, the most difficult one to measure reliably is the in-chain conductivity (resistivity) since any spurious voltage
drop perpendicular to the chain direction can lead to widely erroneous resistivity values and an incorrect temperature dependence. While the Montgomery technique is a reliable method for
determining the inter-chain components, in our experience, it is not an effective means of isolating the small in-chain resistivity since its applicability relies on the ability to short out
the voltage drop in the third orthogonal direction which often cannot be guaranteed. In our experiments therefore, we choose to mount our electrical contacts in such a way that the voltage
drop in both orthogonal directions is shorted out rendering the current flow between voltage contacts strictly one-dimensional. This is achieved by having current pads that completely cover
the ends of the (bar-shaped) sample, using either conductive paint or evaporated gold strips, and voltage pads that are painted (or evaporated) across the entire width of the sample as well
as down the sides. A detailed description of our determination of $\rho_b(T)$ has already been published elsewhere, and we direct the interested reader to the Supplementary Information
section of Ref. \cite{Wakeham} for further details. Typical dimensions of our $\rho_b(T)$ crystals were 750 x 150 x 20 $\mu$m$^3$ and the excitation currents in this case ranged between
10 and 100 $\mu$A.

We note that previous resistivity data reported in the literature and performed using the Montgomery method have a much lower electrical anisotropy than ours, with $\rho_b$($T$) displaying a
sub-linear temperature dependence at high $T$ reminiscent of the $a$- or $c$-axis resistivities. Of course, a low resistivity value (for current along the chains) does not automatically
guarantee that one has isolated the in-chain current response, since current paths can sometimes be such that lower (and even negative) resistivity values are obtained artificially. However,
a strong indicator that we are measuring the intrinsic $b$-axis response is the reproducibility of our data for a number of different samples. Again, we refer the interested reader to the
Supplementary Information of Ref. \cite{Wakeham} for more detail.  In total, we obtained $\rho_b(T)$ curves for over 30 single crystals to allow us to better identify the intrinsic $T$-dependence of the in-chain resistivity and to determine its absolute magnitude. It is important to acknowledge that the resistivity anisotropy we obtain is not only in good agreement with measured $H_{c2}$ anisotropy, but also with that measured independently by optical conductivity \cite{Choi04}.

Finally, note that the $\rho_b(T)$ curve displayed in Figure 1 in the main manuscript and the one we believe to be the most representative of the intrinsic $b$-axis resistivity,
has a lower limit of only 4.2 K. This is because the data were taken in a $^4$He system that could not be pumped. Within a single batch however, the $T_c$ values (and the size of their resistive upturns) tend to vary by less than 10\% (40\%) respectively. Moreover, as shown in a later section of the Supplementary Information, we observed a large specific heat anomaly at $T_c$ = 2.2~K in one of these crystals, confirming that the superconductivity is indeed bulk and three-dimensional.

\section{Determination of $H_{c2}$}

In Figures 3 and 4 of the Letter, we determine $H_{c2}$ using the maximum in the temperature or field derivative of $\rho_c$. In this section, we consider the effect of adopting different
criteria in the determination of the superconducting anisotropy in \LMO. Figure S1 shows a typical $\rho_c(H)$ sweep for {\bf H}$\| b$. The temperature of the sweep is $T$ = 0.81 K. As
might be expected, the maximum in d$\rho_c$/d$T$ gives a near-identical $H_{c2}$ value to that determined by the midpoint of the transition. The 10\% criterion, i.e. the field at which
$\rho_c$ reaches 10\% of its maximum value, is relatively straightforward to identify and can be used to set a lower limit on $H_{c2}(T)$. For the upper limit, the 90\% criterion is much
harder to define due to the broad fluctuation regime (for {\bf H}$\| b$). However, if instead we use the crossing point between the mixed-state and normal-state resistivities for the
definition of our high $H_{c2}$ value (the so-called \lq Junction' criterion of Lee {\it et al.} \cite{Lee02}), as illustrated in Figure S1, we obtain the three sets of curves shown in
Figure S2. Consistent data were obtained on a second crystal, as described in the following section.

As can be seen, the $T$-dependence of $H_{c2}$ is only slightly dependent on the criterion used. More importantly, we find that the anisotropy of the initial slopes is relatively insensitive to the different criteria.  After scaling the $H_{c2}$ values obtained for {\bf H}$\| b$ in the $^3$He system to those obtained in the pumped $^4$He system (due to a slight misalignment of
the applied field, as explained in the main text), we obtain {\bf H}$\| a$:$b$:$c$ d$H_{c2}$/d$T|_{T \leq T_c}$ values of -1.1:-17:-0.4~T/K and -2.4:-30:-0.7~T/K for the 10\% and
crossing-point criteria respectively, corresponding to critical field ratios of 2.8:44:1 and 3.5:44:1 for {\bf H}$\| a$:$b$:$c$ respectively. These compare favorably with the ratio
3:39:1 obtained using the maximum in d$\rho_c$/d$T$, affirming that the anisotropy ratios are robust. Finally the linear extrapolations of the different $H_{c2}(T)$ curves
plotted in Figure S2 provide the following estimates of the uncertainty in our $H_{c2}$(0) values: $H_{c2}^a(0)$ = 5.0 $\pm$ 0.5 T, $H_{c2}^b(0)$ = 15.2 $\pm$ 0.2 T, $H_{c2}^c(0)$ = 1.2
$\pm$ 0.2 T.

\section{$H_{c2}$ of a single crystal with a lower $T_c$}

\begin{figure}
	\centering
		\includegraphics[width=8.0cm,keepaspectratio=true]{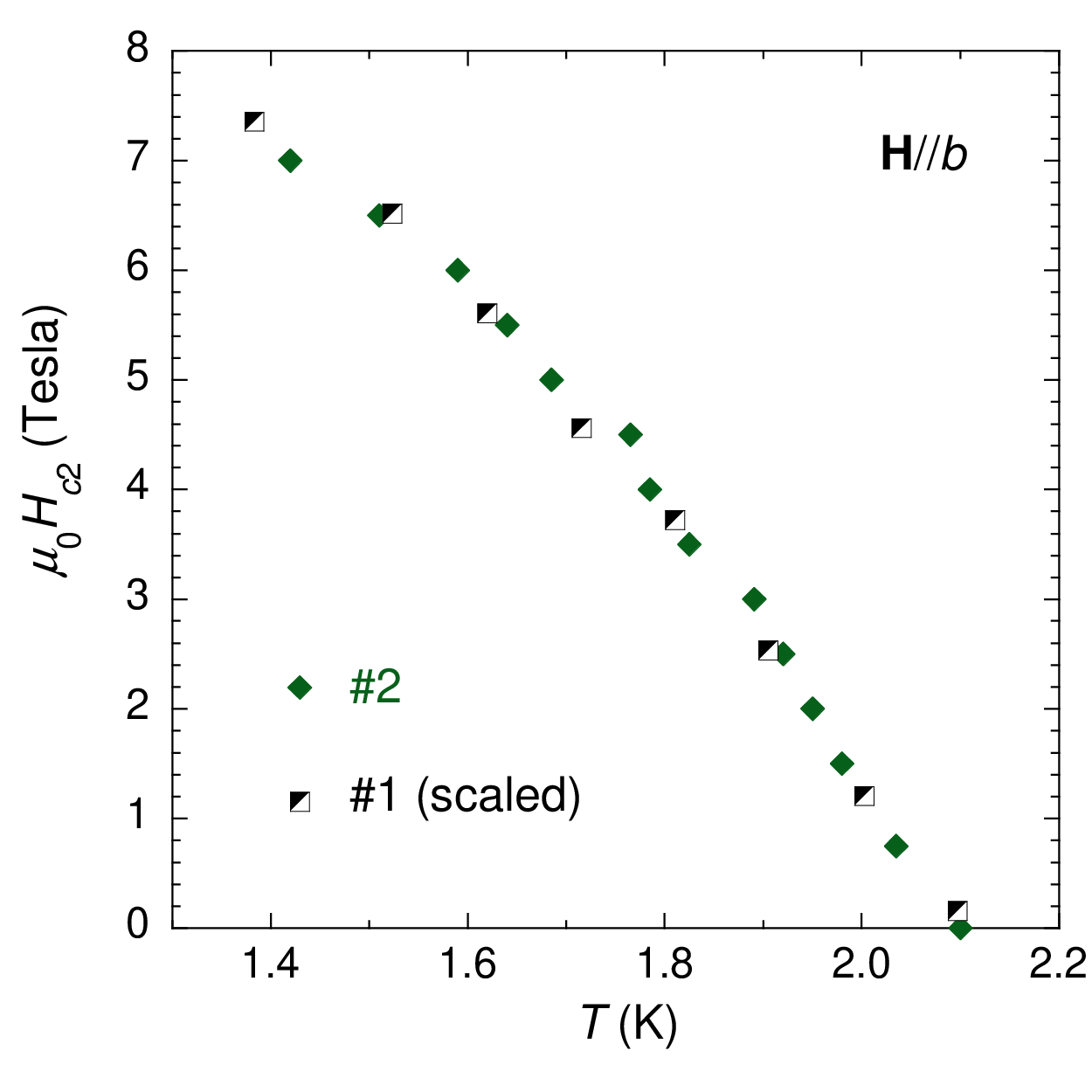}
		\caption{The temperature dependence of the upper critical field $H_{c2}(T)$ for a second crystal ($T_c$ = 2.1 K) and the same crystal described in the main manuscript.}
	\label{fig:S3}
\end{figure}

In our continuing search for field-induced superconductivity in this new batch of crystals, we found similar $H_{c2}$(0) values (i.e. $>$  13 Tesla) in at least two other samples. Figure S3 shows the results of a more detailed study of the $H_{c2}(T)$ profile for {\bf H}$\| b$ for a second sample with a slightly lower $T_c$ (= 2.1 K) and a correspondingly lower $H_{c2}$(0) (= 13 T). Superimposed on this curve are data for the same crystal described in the main manuscript (these are the same data as shown in Figure 4), scaled by a factor of 0.92. Thus, for a 5\% reduction in $T_c$, there is a corresponding 8\% reduction in $H_{c2}$(0). The $T$-dependence of the upper critical field however is essentially identical in the two samples.

\section{Estimating the Pauli limit from heat capacity and magnetic susceptibility measurements}

\begin{figure}
	\centering
		\includegraphics[width=8.0cm,keepaspectratio=true]{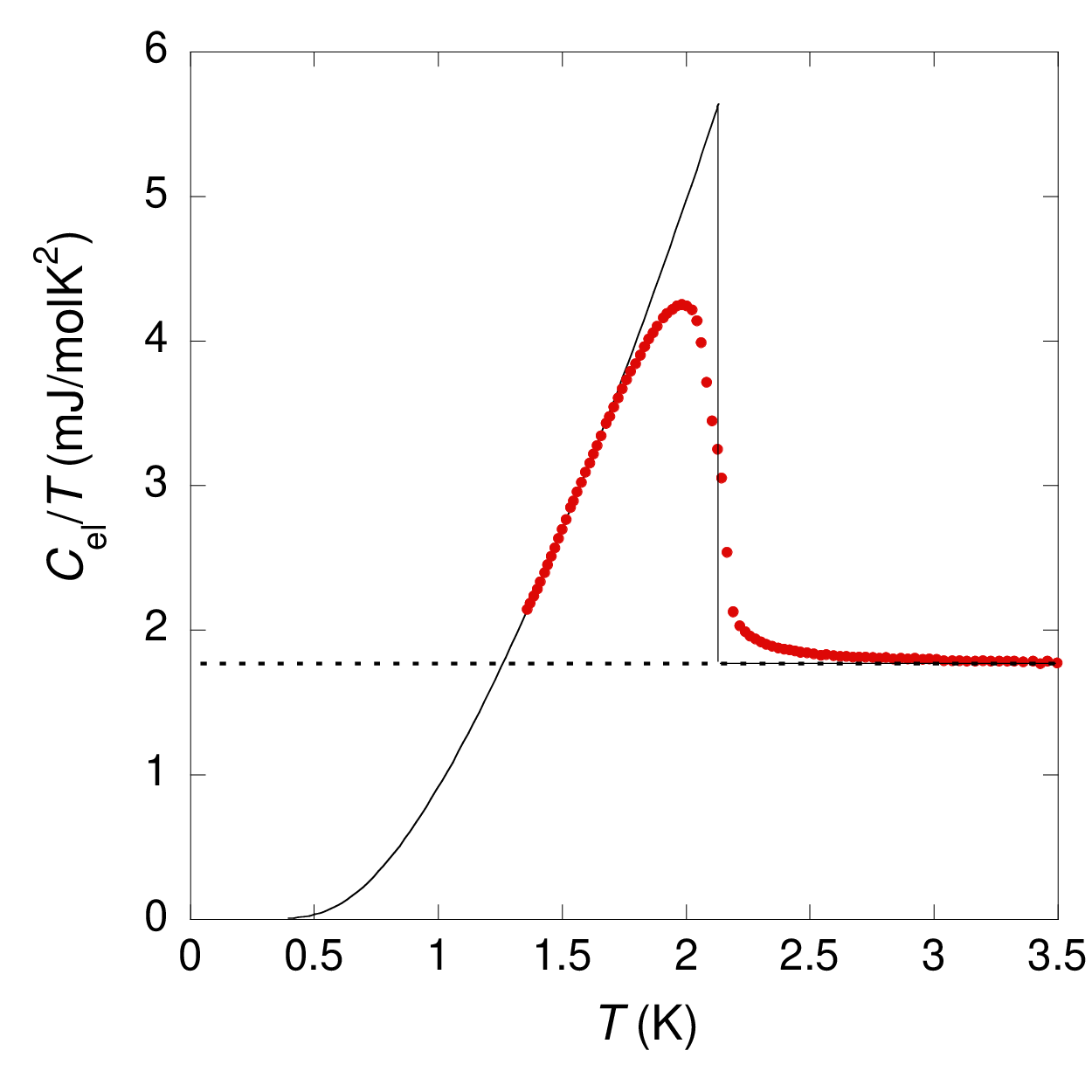}
		\caption{The low-temperature specific heat of a Li$_{0.9}$Mo$_6$O$_{17}$ single crystal, $\gamma$, divided by temperature. The solid line is a fit to the alpha-model for a strongly-coupled superconductor as described in the text. The dashed line is a guide to the eye.}
	\label{fig:S4}
\end{figure}

Figure S4 shows heat capacity data, divided by temperature, for a single crystal from the same batch as the crystal described in the main text. Measurements were conducted using a long-relaxation calorimeter \cite{Taylor07} on a sample with mass $\sim$ 4.4\,mg.  A large specific heat anomaly, associated with the superconducting transition, is observed centered at $T$ = 2.15 K, in excellent agreement with the resistive transition reported in Figure 1. The size of the jump in $C$ at $T_c$, is far in excess of that expected for weak-coupling superconductivity ($\Delta C/\gamma T_c$=2.07), so we have fitted the data using the strong-coupling alpha model \cite{Padamsee73}. Here the temperature dependence of the energy gap is assumed to be the same as in the weak-coupling $s$-wave BCS model, but the absolute value is multiplied by a constant $\alpha$ to account for the enhanced value of $\Delta/T_c$ due to strong coupling. This model has been found to account well for data, both for conventional strong-coupling superconductors \cite{Padamsee73}, as well as more exotic materials such as MgB$_2$ \cite{Bouquet01} and the organic superconductors \cite{Taylor07}. The fit gives the only free parameter in this model,  $\alpha$ = 1.6. By integrating the fitted curve, we get a superconducting condensation energy of $U_c$ = 2.2 mJ/mol ($\sim$ 11 J/m$^3$). In many different electron-phonon coupled superconductors there is a universal relation between  $\Delta C/\gamma T_c$ and the ratio of $T_c$ to the average phonon-frequency $\omega_{\ln}$, which is approximately given by \cite{Carbotte90}
\[
\frac{\Delta C}{\gamma T_c} = 1.53\left[1+53\left(\frac{T_c}{\omega_{\ln}}\right)^2\ln\left(\frac{\omega_{\ln}}{3T_c}\right)\right].
\]
This gives, $\omega_{\ln}/T_c$ = 0.075, which is intermittent between Nb and Pb.  Using this we can solve the McMillan equation
\[
1.2 T_c=\omega_{\ln} \exp\left[\frac{-1.04(1+\lambda)}{\lambda-\mu^*(1+0.62\lambda)}\right],
\]
to obtain a value for the coupling constant $\lambda=1.23$ for $\mu^*=0.15$ ($\lambda=1.12$ for $\mu^*=0.12$).

\begin{figure}
	\centering
		\includegraphics[width=8.0cm,keepaspectratio=true]{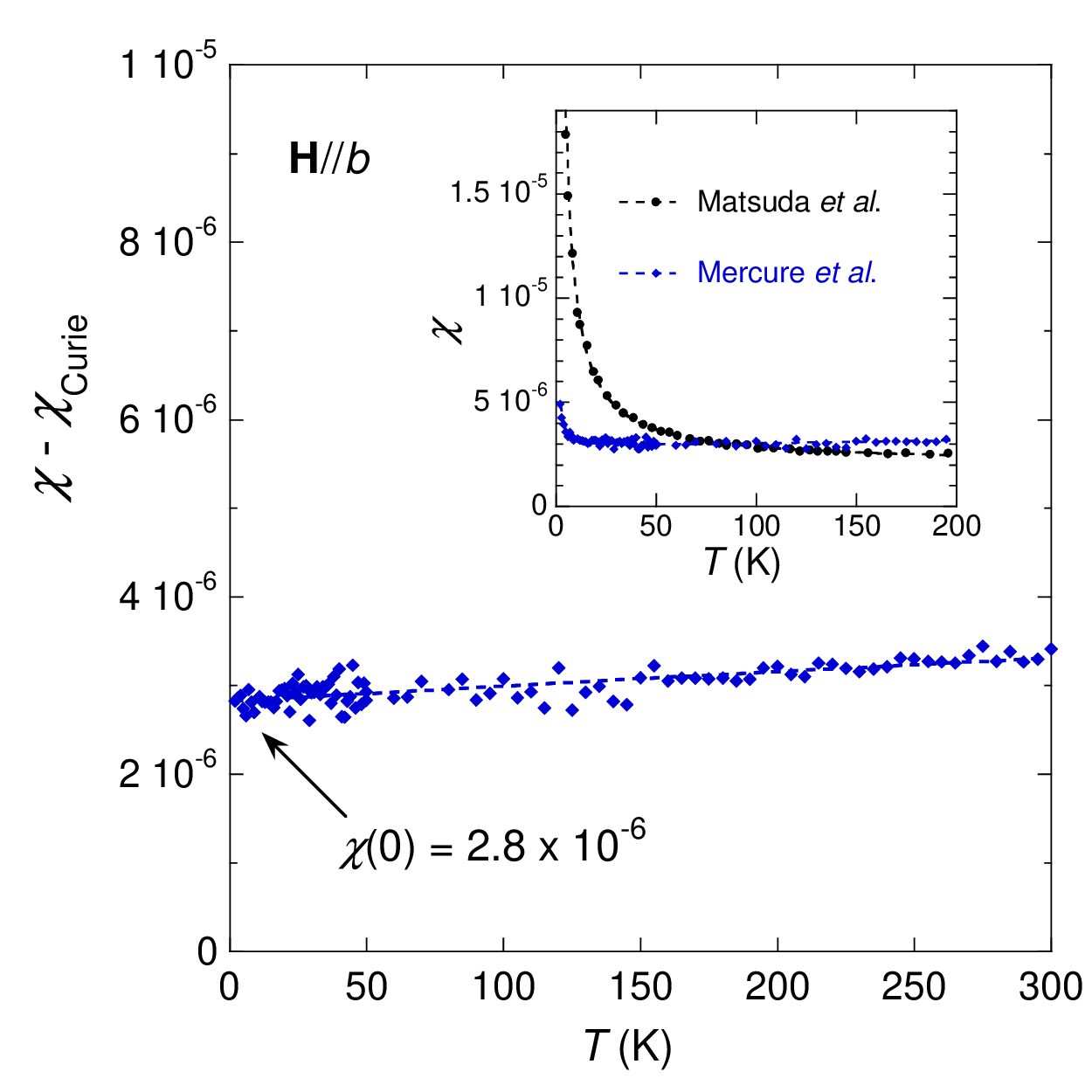}
		\caption{Main panel: Residual magnetic susceptibility data for Li$_{0.9}$Mo$_6$O$_{17}$ obtained from subtraction of the 1/$T$ Curie term. The dashed line is a guide to the eye. Inset: Raw susceptibility data for our Li$_{0.9}$Mo$_6$O$_{17}$ single crystal (solid symbols) and for a Li$_{0.9}$Mo$_6$O$_{17}$ single crystal reported by Matsuda {\it et al.} \cite{Matsuda86} (open circles).}
	\label{fig:S5}
\end{figure}

The inset to Figure S5 shows raw magnetic susceptibility $\chi(T)$ data for another single crystal from the same batch measured between 2 K and 200 K in a magnetic field of 5 Tesla applied along the $b$-axis. The same crystal also shows a superconducting transition in lower fields at $T_c$ = 2.0 K. Plotted alongside this data set are earlier $\chi(T)$ data reported by Matsuda {\it et al.} back in 1986 \cite{Matsuda86}. While the high temperature values are similar, the data of Matsuda {\it et al.} show a much larger enhancement of low $T$. This enhancement was attributed to an extrinsic Curie paramagnetic term caused by a concentration of 0.15 x 10$^{-3}$ localized ($S$ = 1/2) moments per conduction electron. In our crystals, the enhancement is a factor of 20 smaller. Subtracting this small 1/$T$ term from the raw data, we obtain the curve plotted in the main panel of Figure S5. The residual susceptibility data show a very weak $T$-dependence, extrapolating to a zero-temperature value $\chi(0)$ = 2.8 x 10$^{-6}$. The ionic diamagnetism $\chi_{dia}$ and the orbital van Vleck contribution $\chi_{vv}$ to $\chi(T)$ are believed to be comparable in Li$_{0.9}$Mo$_6$O$_{17}$, with $\chi_{dia}$ having the slightly larger magnitude \cite{Matsuda86}. This implies that the residual susceptibility data is a lower estimate of the full Pauli paramagnetic susceptibility term $\chi_P$. We can also obtain an estimate for $\chi_P$ from the measured Sommerfeld constant $\gamma(0)$ (= 1.6 mJ/mol.K$^2$), assuming a Wilson ratio $R_W$ = 2 (that is consistent with most strongly correlated metals as well as Luttinger liquids with repulsive interactions). From these considerations, we obtain $\chi_P$ = 3.0 x 10$^{-6}$, in good agreement with the measured value.

Combining our estimates for $U_c$ and $\chi_P$, we obtain an estimate for the Pauli limit of $B_P$ = (2$\mu_0 U_c$/$\chi_P$)$^{0.5}$ = 3.1 Tesla, i.e. lower than the limit estimated simply from the value of $T_c$. This estimate can be revised upward however if one considers the effects of strong coupling. According to Figure 82 of Carbotte's review of strong-coupling superconductivity \cite{Carbotte90}, given the slope of $H_{c2}(T)$ near $T_c$ ($\sim$ 14 T/K, as measured in the $^3$He system) and our estimate of $\lambda$, one expects a corresponding Pauli limiting field of 7.3 Tesla in the clean limit. However, the actual value of $H_{c2}$(0) in our crystals ($>$ 15 T) is more than a factor of 2 larger than this strong-coupling expectation.

\end{document}